\documentclass[lettersize,journal]{IEEEtran}
\usepackage{amsmath,amsfonts}
\usepackage{algorithmic}
\usepackage{algorithm}
\usepackage{array}
\usepackage[caption=false,font=normalsize,labelfont=sf,textfont=sf]{subfig}
\usepackage{textcomp}
\usepackage{stfloats}
\usepackage{url}
\usepackage{verbatim}
\usepackage{graphicx}
\usepackage{booktabs}
\usepackage{cite}
\usepackage{hyperref}

\usepackage[final]{changes} 
	\definechangesauthor[name={}, color=orange]{S.} 
\usepackage{color}
\usepackage{hyperref}

\hypersetup{hypertex=true,
            colorlinks=true,
            linkcolor=blue,
            anchorcolor=blue,
            citecolor=blue}
\begin{document}

\title{OS-FPI: A Coarse-to-Fine One-Stream Network for UAV Geo-Localization}


\author{
	 Jiahao Chen, Enhui Zheng, Ming Dai, Yifu Chen, Yusheng Lu,
	\thanks{Jiahao Chen, Enhui Zheng, Ming Dai, Yifu Chen, Yusheng Lu, are with the Unmanned System Application Technology Research Institute, China Jiliang University, Hangzhou, 310018, China. (email:p21010854010@cjlu.edu.cn; ehzheng@cjlu.edu.cn; S20010802003@cjlu.edu.cn; p22010854011@cjlu.edu.cn, p21010854069@cjlu.edu.cn;).Enhui Zheng is the Corresponding Author.}
}

\markboth{Journal of \LaTeX\ Class Files,~Vol.~14, No.~8, August~2021}%
{Shell \MakeLowercase{\textit{et al.}}: A Sample Article Using IEEEtran.cls for IEEE Journals}


\maketitle

\begin{abstract}
The geo-localization and navigation technology of unmanned aerial vehicles (UAVs) in denied environments is currently a prominent research area. Prior approaches mainly employed a two-stream network with non-shared weights to extract features from UAV and satellite images separately, followed by related modeling to obtain the response map. However, the two-stream network extracts UAV and satellite features independently. This approach significantly affects the efficiency of feature extraction and increases the computational load. To address these issues, we propose a novel coarse-to-fine one-stream network (OS-FPI). Our approach allows information exchange between UAV and satellite features during early image feature extraction. To improve the model's performance, the framework retains feature maps generated at different stages of the feature extraction process for the feature fusion network, and establishes additional connections between UAV and satellite feature maps in the feature fusion network. Additionally, the framework introduces offset prediction to further refine and optimize the model's prediction results based on the classification tasks. \replaced{Our proposed model, boasts a similar inference speed to FPI while significantly reducing the number of parameters. It can achieve better performance with fewer parameters under the same conditions.}{As a one-stream network, our proposed model not only improves inference speed, but also reduces the number of parameters.} Moreover, it achieves state-of-the-art performance on the UL14 dataset. Compared to previous models, our model achieved a significant 10.92-point improvement on the RDS metric, reaching 76.25. Furthermore, its performance in meter-level localization accuracy is impressive, with 182.62\% improvement in 3-meter accuracy, 164.17\% improvement in 5-meter accuracy, and 137.43\% improvement in 10-meter accuracy.

\end{abstract}


\section{Introduction}
\IEEEPARstart{W}{ith} the ever-advancing remote sensing and satellite technology, obtaining high-resolution satellite imagery has become increasingly feasible. These images now have a global reach, spanning rural and urban areas alike. Researchers can analyse and process remotely sensed images to get the key data they need. The continuous progress of remote sensing technology cannot be separated from two key technologies, one is the updating and iteration of sensors, in addition to visible light, infrared sensors as well as Synthetic Aperture Radar (SAR) and other advanced equipment also provide a strong impetus for the development of remote sensing technology\cite{bib41,bib42,bib46,bib48,bib50}. Another is the continuing breakthrough in the field of computer vision, and the development of small object detection, object tracking, image alignment, image retrieval and other technologies in this neighbourhood has received great attention\cite{bib40,bib43,bib45,bib47,bib49}. Cross-view Geo-Localization technology is also one of them.

Cross-view geolocation refers to determining the location information of the current query image by comparing images containing location information in the retrieval database. Unmanned aerial vehicles (UAVs) heavily rely on GPS data provided by satellite signals while in operation. However, both civilian and military sectors experience a significant number of flight accidents due to the loss of satellite signals. This is an ongoing issue that needs to be addressed to ensure safer UAV operations. The utilization of imaging methods for positioning and navigation of UAVs in challenging environments will have significant implications in the coming years.

The ongoing advancement in computer vision technologies, such as object detection, image retrieval, and object tracking, offer the potential of relying solely on visual information for UAV geolocation and navigation tasks. Current cross-view geolocation technology for UAVs is mainly realized through two approaches: image retrieval \cite{bib10,bib12,bib17,bib18} and the method of finding points with images \cite{bib19,bib20}.

The method of image retrieval is mainly through supervised learning, so that the features of the same area of the picture are constantly approaching, and the features of different areas are constantly moving away, so as to achieve the matching between images. In previous work, researchers have done a lot of related work, including matching UAV images with satellite images, and matching UAV images with street view images, etc. Nonetheless, certain elements hinder the enhancement of positioning precision in image retrieval. On the one hand, the images in the database cannot cover the entire area. The larger the area covered, the more data the computer needs to hold, and it also takes more inference time. On the other hand, since it is difficult to ensure that the images in the database and the image to be queried are centrally aligned, there will be a great deviation in positioning accuracy. Due to the problems mentioned above, we need to prepare a large-scale database in advance when applying the image retrieval method in the actual flight process. At the same time, the query image needs to calculate the similarity with each image in the database, which is tedious work. It is very poor for UAV positioning and navigation tasks.

In order to solve the problems of poor positioning accuracy and time-consuming application, a new method of finding points with images was proposed \cite{bib19,bib20}. It borrows from the field of object tracking by modeling the relationship between UAV images in vertical view and satellite images to obtain a response map. The point with the largest value in the heat map is where the model predicts the center of the UAV image to be in the satellite image. The method uses a two-stream network in the feature extraction stage, that is, two backbones that do not share weights, for extracting feature maps of UAV images and satellite images respectively, followed by modeling the relationship between UAV feature maps and satellite feature maps. Thus, the entire network structure can be divided into two parts: feature extraction and information interaction. It is worth mentioning that the two parts are independent of each other.

Although the method of finding points with images has achieved some results, there are still several problems with such a two-stream network: 1) When the model extracts the features of UAV and satellite images in the early stage, there is no information interaction between the two, which makes it difficult for the model to extract features that are effective for this task. 2)In the previous two-stream network, group convolution was often used to model the relationship between UAV feature maps and satellite feature maps. However, the receptive field of convolution is limited and lacks global modeling capabilities. For such a fine-grained task, the context information of the picture will have a key impact on the positioning effect. 3) Using a two-stream network and complex relational modeling methods will bring more parameters and computational consumption, which will greatly reduce the speed of model reasoning.

\begin{figure*}[!t]
\centering
\includegraphics[width=7 in]{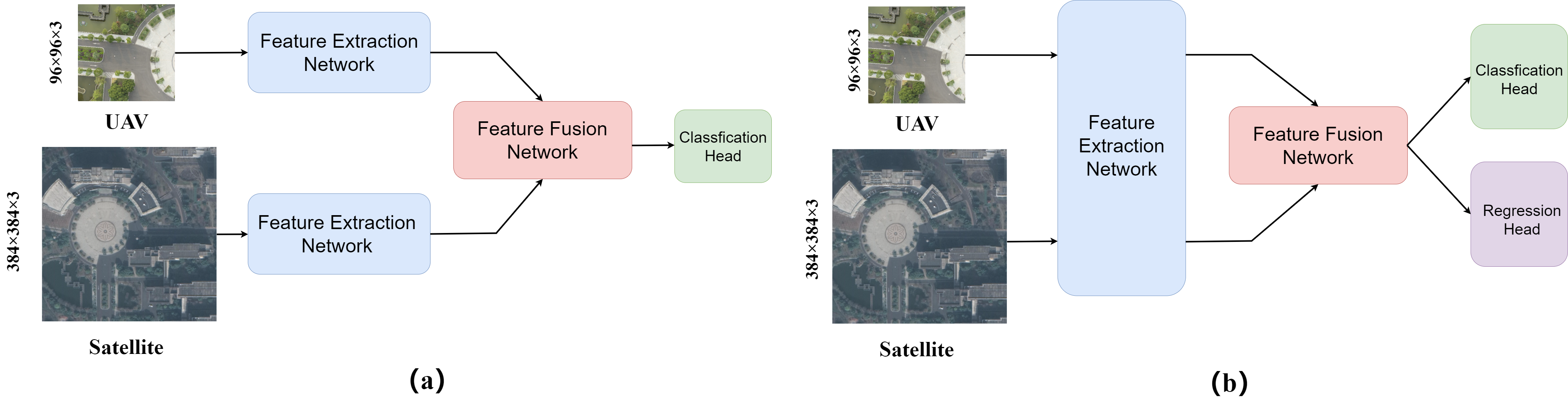}
\caption{The comparison of one-stream and two-stream networks: (a) The conventional two-stream network, which uses two feature extraction networks that do not share weights to extract features from UAV images and satellite images separately, and then constructs a link between UAV images and satellite images through a feature fusion network. (b) In this paper, we propose a one-stream network that establishes a bridge between UAV and satellite images through a flexible attention mechanism in the feature extraction module, and strengthens the connection between them through a feature fusion network. Additionally, we introduce a regression task, in addition to the classification task, for offset prediction.}
\label{fig_1}

\end{figure*}

To solve the above problems, we propose a one-stream network. Figure \ref{fig_1} is a comparison diagram of a one-stream network and a two-stream network. Our proposed network integrates traditional feature extraction and relational modeling by utilizing a shared backbone to process both UAV and satellite images. During feature extraction, we leverage the flexibility of the Transformer mechanism to establish a channel for information interaction between UAV and satellite features. This method of joint feature extraction and information interaction has the following characteristics: 1) At the early stage of feature extraction, our model can determine the relevant features to retain, significantly enhancing the efficiency of the process and minimizing the loss of target information. 2) Using the Transformer mechanism to create a global connection between UAV and satellite images facilitates information interaction between the two, resulting in improved performance. 3) Feature maps were saved for each stage in order to subsequently establish more interaction between UAV and satellite features.

In addition to the backbone, we have also improved the original feature fusion network. To prevent a reduction in localization accuracy resulting from the decreased resolution of the prediction map, we introduced a feature pyramid structure into our model after the initial extraction of UAV and satellite images. It is worth mentioning that in the one-stream network, only the satellite feature map uses the feature pyramid structure. After that, we also introduced atrous convolution to improve the model's ability to obtain context information. The shallow feature map retains more texture information than the deep feature map. After using the shallow UAV feature map to model the relationship with current features again, we surprisingly found that the method can further improve the localization performance of the model, so we added more links between the UAV feature map and the satellite feature map in the feature fusion network. Finally, we improved the head. In previous methods, the point with the largest value in the prediction map was used as the final prediction result. This is a pixel-level classification task. To achieve a more fine-grained localization effect, we introduce offset prediction, which adjusts the predicted results on the basis of classification. This method can reduce the problem of inaccurate localization due to reduced resolution, and can also adjust the results beyond the unit pixel range. This enhances the network's positioning capabilities.

The following is a summary of our work:
\begin{enumerate}
    \item We propose a novel end-to-end network framework that introduces cross-attention operations. While realizing early feature extraction of pictures, a bridge of information communication is established between UAV pictures and satellite pictures. And the introduction of the SRA structure effectively reduces the computational overhead and improves the speed of network reasoning.
    \item In the task of finding points with images, we proposed a new method of joint training of classification and regression, and developed a new loss function on this basis. This is a coarse-to-fine prediction method, and we introduce offset predictions on top of classification predictions, which, to the best of our knowledge, have not been explored in previous studies.
    \item Our proposed model achieves state-of-the-art performance on the UL14 dataset, surpassing the previous model by improving the RDS metric by 10.92 points to 76.25. The improvements in meter-level positioning accuracy are equally impressive, with a 182.96\% improvement (from 12.49\% to 22.81\%), a 164.17\% improvement (from 26.99\% to 44.31\%), and a 137.43\% improvement (from 52.62\% to 72.32\%) in the positioning accuracy within 3 meters, 5 meters, and 10 meters, respectively.
\end{enumerate}

\section{Related Work}

Early cross-view geolocation technology was mainly implemented by image retrieval technology. That is, the query image was used to find the most relevant images in a database containing location information, so as to obtain the location information of the query image. These tasks include matching ground images to other ground images \cite{bib1,bib2,bib3,bib4,bib5,bib6,bib7}, matching ground images to aerial images, etc. \cite{bib8,bib9,bib10,bib11,bib12}. For example, in \cite{bib13} the authors first proposed the use of convolutional neural networks to solve the cross-view geolocation problem. In \cite{bib8}, the authors proposed a novel convolutional neural network that aims to associate semantic information between aerial images and ground-based street images of the same region. In \cite{bib14}, the authors integrated a transformer structure into the network and employed a non-uniform cropping strategy to eliminate a considerable amount of irrelevant information and reduce computational costs. 

Although the above approach has achieved some success, since there are often significant feature variations between images from different viewpoints, learning the same features for different viewpoints becomes a challenging task for cross-view geo-localization. To address this challenge, \cite{bib15} proposed a method of aligning aerial images with satellite images using polar coordinate transformation to bridge the differences between the two. \cite{bib16} converting Street View Images to UAV Images by Generative Adversarial Networks to Reduce Matching Inaccuracies Due to Dramatic viewpoints Changes. \cite{bib17} introduced GeoNet, which utilizes a spatial hierarchical structure for modeling to learn viewpoint-invariant features in cross-view images. The above methods all utilize traditional techniques to align multi-source data. However, in \cite{bib38}, the authors propose the use of weight sharing to extract features from two images at the same time. This approach aims to fully leverage the relationships present within multi-source data. Additionally, the method introduces edge feature information and salient features based on an attention mechanism to enhance the matching performance. These ideas presented in \cite{bib38} also provide inspiration for the current paper. 

With the continuous development of UAV technology and satellite remote sensing technology, the work between UAV domain and satellite domain has become a hot spot. \cite{bib10} introduced a novel dataset, University-1652, which comprises data from three platforms: ground, UAV, and satellite. They also introduced a new task of UAV visual localization and navigation. To bridge the view gap between the \begin{math}45^{\circ}\end{math} oblique view and the satellite image, \cite{bib18} used an end-to-end cross-view matching method combining a cross-view synthesis module and a geolocation module to reduce the learning burden of cross-view matching by converting the oblique view UAV images to satellite images through perspective transformation and conditional generation adversarial networks, thus improving the model performance.

Currently, research on the utilization of image retrieval for visual Geo-localization is rapidly expanding. In this regard, the establishment of a benchmarking framework is also considered crucial. \cite{bib39} introduces a framework, which makes the construction of the model training and testing become more standardized and flow, the user can flexibly train and test this task. This framework not only simplifies the development and evaluation processes but also facilitates the reproducibility and comparison of results across different studies.

However, cross-view geolocation by means of image retrieval has heavily relied on the assumption that the database contains images aligned with the query image. This does not apply in the real scenario. What do we want? Given a picture of a UAV in any area, the current location of the UAV can be found in the database. To this end, \cite{bib19} proposed a new end-to-end method of finding points with images and a new UL14 dataset, where the authors used a two-stream network without shared weights to extract the vertical view of the UAV image and the satellite image respectively, after which the response map was obtained by relational modeling, and the point with the largest response value in the map was the current position of the UAV image predicted by the model. This end-to-end approach provides a completely new direction for the development of UAV visual localization, and the authors also propose an MA metric to quantify error, using meters as the unit of measurement for error. On the basis of FPI\cite{bib19}, in order to alleviate the multi-scale problem in the task of finding points with images, the author proposed the WAMF module \cite{bib20}. And the final output response map is restored to the original satellite map size, thus reducing the problem of inaccurate positioning due to the small resolution of the prediction map and further improving the positioning accuracy of the model. However, both image retrieval and finding points with images in the early stages of feature extraction are carried out using siamese networks with non-shared weights, resulting in disconnected feature extraction between the images from different branches. This significantly hinders the efficiency of feature extraction.

As part of computer vision tasks, the visual object tracking task has also made great progress in recent years. From correlation filtering methods to current deep learning- based methods, the most representative network is Siamfc \cite{bib21}. As the initial installment in the Siamese series, it established the groundwork for following visual object tracking tasks. After that, SiamRPN\cite{bib22}, SiamRPN++\cite{bib23} and siammask\cite{bib24} added tasks such as RPN structure and semantic segmentation to the network to further improve network performance. Visual object tracking can be broadly divided into two parts: a backbone for extracting generic features, and an information interaction network for relationship modeling. Previous research has extensively explored Siamese network-based methods for various tasks \cite{bib21,bib22,bib23,bib24,bib25,bib26,bib27,bib28}. Unlike cross-view geolocation where images come from different views, the object tracking task uses the first frame of the image as a template image. Therefore, a Siamese network with shared weights is used in object tracking. That is, the template image and the search image use the backbone with the same parameters to extract early features. Relational modeling is performed on it later. In recent years, there have been significant advancements in the field of object tracking. Instead of using Siamese networks with shared weights in mixformer \cite{bib29} and os-track \cite{bib30}, feature extraction is combined with information interaction. This approach further improves the efficiency of feature extraction. Drawing on the latest object tracking framework, we present OS-FPI, which outperforms previous models with fewer parameters, achieving superior performance.

\begin{figure*}[!t]
	\centering
	\includegraphics[width=7 in]{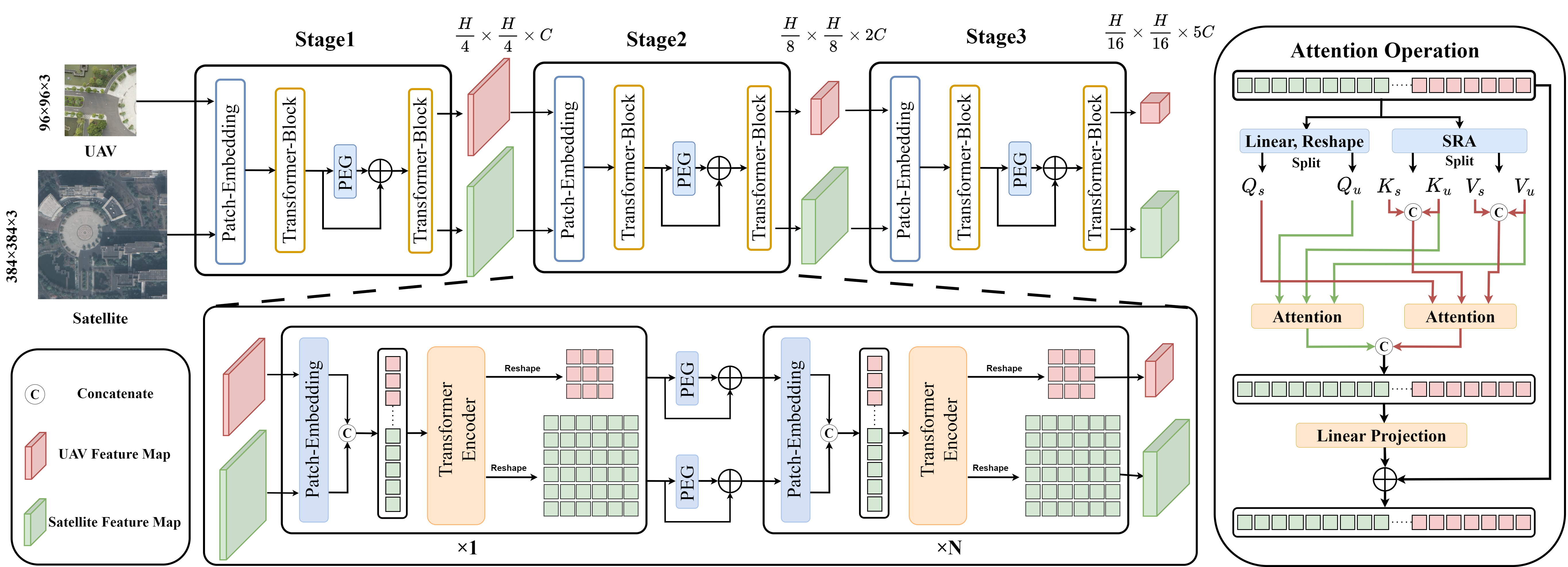}
	\caption{The entire backbone is divided into three stages, and a Position Encoding Generator (PEG) is added after the first Transformer encoder of each stage to replace the absolute position encoding. Additionally, the feature maps of the three stages are gradually downscaled in a pyramid structure. The number of channels for the feature maps in the three stages are 64, 128, and 320, respectively. The right side of the figure is the key part of joint feature extraction and relationship modeling, which is the Attention operation in the Transformer encoder.}
	\label{fig_2}
\end{figure*}

\section{Method}
This chapter introduces our proposed end-to-end framework, OS-FPI, which is the first application of joint feature extraction and information interaction method to cross-view geo-localization and navigation tasks to the best of our knowledge. We first introduced the overall structure of OS-FPI in Section \ref{3.1}. Later, in Section \ref{3.2}, we explain how the OS-FPI framework incorporates initial feature extraction and information integration. In our proposed method, UAV images and satellite images are fed into the same backbone. In Section \ref{3.3}, we present the feature fusion network of the framework, which seeks to create additional links between UAV and satellite images. Finally, in Section \ref{3.4}, we introduce offset prediction, which is used in the model to adjust the results of classification, a more fine-grained approach that can further improve the accuracy of model localization.

\subsection{Overall Architecture}
\label{3.1}
Given a set of UAV images and satellite images, our goal is to find the location of the center of the UAV images in the satellite images. As shown in Figure \ref{fig_2}, UAV and satellite images are first fed into a joint feature extraction and information interaction backbone , and the connection between them is established while extracting the features of UAV images and satellite images, we refer to this backbone as OS-PCPVT. The Transformer's global modeling property enables us to establish connections between different features. As shown in Figure \ref{fig_2}, the backbone is composed of three stages, each of which generates two feature maps of varying sizes, corresponding to the UAV feature map and the satellite feature map, respectively. Upon completing the three stages of the backbone, three distinct scales of UAV and satellite feature maps are generated. These feature maps are utilized in the feature fusion network, as well as to establish additional connections between UAV and satellite features. Finally, we introduced offset prediction in the model to further adjust and optimize the prediction results of the model to reduce localization error. Next, we will elaborate on the structure of the model.

\subsection{Feature Extraction Network}
\label{3.2}
In this section, we introduce the proposed OS-PCPVT network for joint feature extraction and relationship modeling. As shown in Figure \ref{fig_1}, the previous method of finding points with images extracts the features of UAV and satellite images, respectively, through two backbones that do not share weights. After this, the information interaction between UAV and satellite images is achieved by simple group convolution or thick multi-scale fusion methods. Therefore, in the early stage of feature extraction, there is no communication between the UAV branch and the satellite branch, and the two-stream network also brings more computational pressure to the model. The baseline for the model is the Twins-PCPVT-S\cite{bib37}, on which we have made a number of improvements, particularly in the Transformer Encoding section. The method proposed in this paper builds a bridge to communicate information between UAV images and satellite images while performing image feature extraction, thus improving the efficiency of feature extraction.

As shown in Figure \ref{fig_2}, the input of OS-FPI is a UAV image with a size of \begin{math}H_{z} \times W_{z} \times 3\end{math} and a satellite image with a size of \begin{math}H_{x} \times W_{x} \times 3 \end{math}. First, we divide them into \begin{math}\frac{H_{z} \times W_{z}}{4^2}\end{math} and \begin{math}\frac{H_{x} \times W_{x}}{4^2}\end{math} patches, respectively, with each patch size \begin{math}4 \times 4 \times 3\end{math}. Then, we feed them into a linear projection, reshaping them into 2D patches whose sizes are \begin{math}
    N_{z}^{2} \times C1
\end{math} and \begin{math}
    N_{x}^{2} \times C1
\end{math}. (\begin{math} N_{z} \end{math} and \begin{math} N_{x} \end{math} represent \begin{math} \frac{H_{z} \times W_{z} }{4^2}\end{math} and \begin{math} \frac{H_{x} \times W_{x} }{4^2}\end{math}, respectively, C1 is the number of channels in the first stage). Finally, we combine them to create an embedded patch with a size of \begin{math}(
    N_{z}^{2}+N_{x}^{2}) \times C1
\end{math}. Next, we input the merged patches into a Transformer encoder, which carries out feature extraction and establishes a channel of communication between the UAV and satellite features for exchanging information. Following the first Transformer encoder in every stage, the original absolute positional encoding is replaced by a PEG (Position Encoding Generator) module\cite{bib31}. 
\added{Conditional Position Encoding (CPE)\cite{bib51,bib52} can be easily implemented through PEG, and CPE can be more flexibly applied to different input sequences while maintaining translation-invariance. These features are crucial for model training and applications.}
After each stage of the OS-PCPVT network, the output is reshaped into a feature map. In the first stage, the size of the UAV feature map is \begin{math}\frac{H_{z}}{4} \times \frac{W_{z}}{4} \times C1 \end{math}, while the size of the satellite feature map is \begin{math}\frac{H_{x}}{4} \times \frac{W_{x}}{4} \times C1 \end{math}. The current output feature maps are then utilized as inputs for the subsequent stage, and this process is repeated. 

After passing through the entire backbone, the model obtains three different scales of UAV feature maps and satellite feature maps, each compressed by a factor of 4, 8, and 16 compared to the original image, respectively. It is worth noting that we removed the last stage of the network, as it compresses the feature maps by a factor of 32, which is unfavorable for this particular task. Our proposed approach integrates feature extraction and information interaction in the backbone, which better captures the correlation between UAV and satellite images. This is a unified method for feature extraction and information interaction. 

The right side of Figure \ref{fig_2} is the attention operation part of the Transformer encoder, which is the core of OS-FPI. The objective is to facilitate information exchange between UAV and satellite features, in order to capture specific information within such features. It is worth mentioning that during the generation of K and V, we introduce the SRA module \cite{bib36}, which reduces the spatial scale of K and V before performing attention calculations. In this way, the computational overhead can be effectively reduced. This enables the processing of larger input feature. After that, the resulting Q, K, and V tensors are partitioned along the spatial dimension into \begin{math} Q_{u}\end{math}, \begin{math} k_{u}\end{math}, and \begin{math} v_{u}\end{math} for the UAV domain, and \begin{math} Q_{s}\end{math}, \begin{math} K_{s}\end{math}, and \begin{math} V_{s}\end{math} for the satellite domain. Finally, we will introduce a cross-attention operation \cite{bib29} between the UAV and the satellite features to realize the information interaction between the two. \added{OS-PCPVT employs asymmetric cross-attention during the operation of the attention mechanism. Figure \ref{fig_2} shows that during attention computation, the UAV features perform self-attention. There are two reasons for this, firstly a complete cross operation will result in more computation and inefficiency. Secondly this method, enhances the information of UAV feature branches.}

\added{OS-PCPVT allows self-attention computation for each sequence, while the combination of feature extraction and relational modelling is achieved by connecting sequences and cross-attention operations. Therefore the method proposed in this paper is fundamentally different from the traditional weight sharing approach. It can achieve the unity of feature extraction and information interaction.}

With the help of the joint feature extraction and information interaction approach, our model reduces more than half of the parameter count and effectively improves the localization performance.

\begin{figure*}[!t]
	\centering
	\includegraphics[width=7 in]{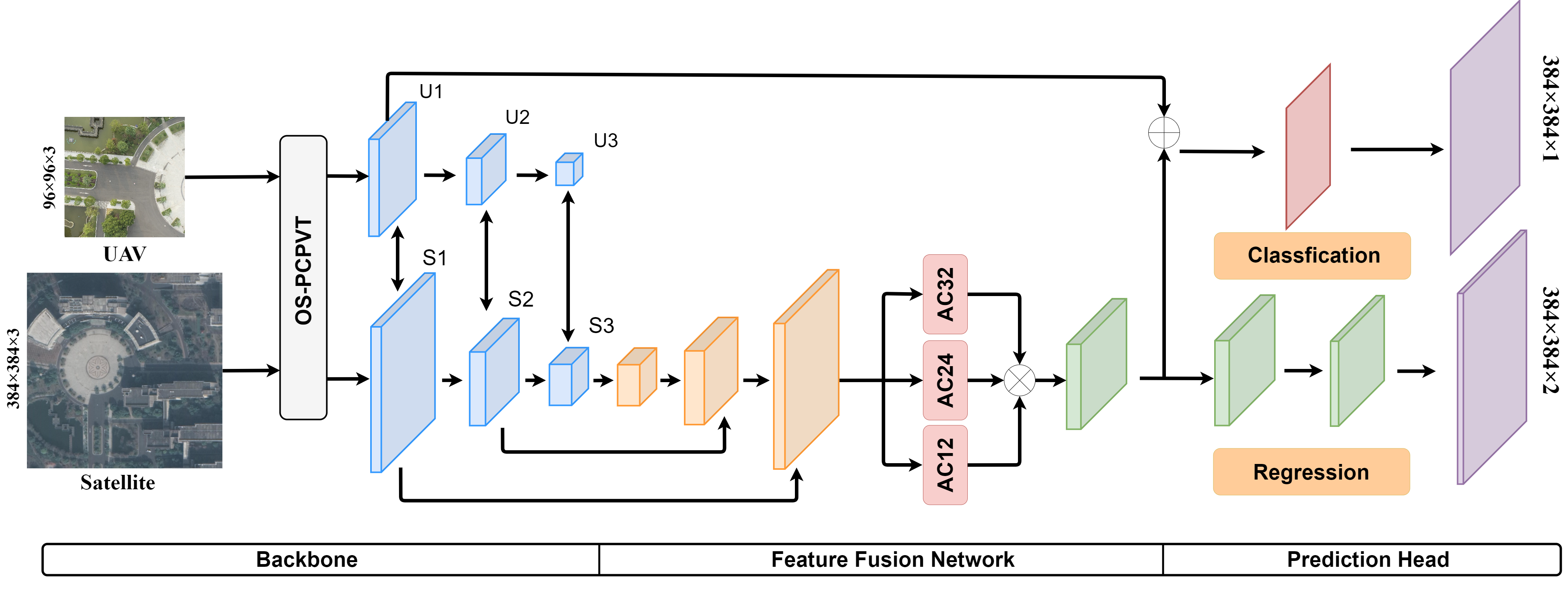}
	\caption{The schematic diagram of feature fusion network. U1, U2, and U3 represent UAV feature maps outputted from three different stages of the backbone, and S1, S2, and S3 represent satellite feature maps outputted from three different stages as well. All of them are derived from the same backbone (OS-PCPVT). This structure fully utilizes the hierarchical architecture of OS-PCPVT, establishing more information exchange between the UAV feature maps and satellite feature maps. Finally, the predicted image is restored to the original satellite image size. Additionally, this framework introduces offset prediction for the first time, further improving the localization performance on top of the classification task.}
	\label{fig_3}
\end{figure*}

\subsection{Feature Fusion Network}
\label{3.3}
\noindent {\bf{Feature Pyramid Structure:}}The task of finding points with images is a fine-grained task, which is very sensitive to changes in pixel compression. As the feature map is compressed more and more, less and less spatial information will be retained in it. Large-scale compression of the output feature map scale will result in significant degradation of localization performance. However, deeper feature maps preserve more abstract semantic information, which is crucial for classification tasks. Therefore, while we restore the feature map output by the model to the scale of the original satellite image, we must also retain more abstract semantic information in the feature map. To this end, we introduce a feature pyramid structure, as shown in Figure \ref{fig_3}. After the backbone, we obtain satellite feature maps at different scales (S1, S2, and S3) from three distinct stages. To merge low-resolution, high-semantic, and high-resolution, low-semantic features among the different feature maps, we employ the feature pyramid structure, which utilizes up-sampling and a lateral connection structure. This results in an output feature map that possesses strong semantic information while maintaining a high resolution. It is worth noting that we did not use the feature pyramid structure for the UAV branch, as modelling the relationship with the satellite branch using a larger feature map would have required a huge amount of computation and a significant amount of inference time.

\noindent {\bf{Atrous Convolution:}}Atrous convolution can effectively expand the receptive field of the convolution kernel and collect more context information at the same time. It is often used in tasks such as semantic segmentation and dense image prediction \cite{bib32,bib33,bib34,bib35}. We believe that the task of finding points with images shares similarities with semantic segmentation. It needs to pay attention to the classification of each pixel and also needs more context information. Therefore, after obtaining the feature map output by the feature pyramid structure, we introduced 3 different atrous convolutions, and their atrous rates are 12, 24, and 32, respectively. As shown in Figure \ref{fig_3}, AC12, AC24, and AC32 represent atrous convolutions with different atrous rates. After extracting the features through different atrous convolutions, they are concatenated along the channel axis and finally fused together using a 3x3 convolutional kernel. In addition, this approach can provide more contextual information without increasing the computational cost, which is crucial for tasks such as image-to-point regression.

\noindent{\bf{Multitasking Training:}}We introduced a regression branch(Offset prediction branch), to the OS-FPI model, which was not previously covered in existing work. After fusing features through different atrous convolutions, the model is split into two branches: a classification branch and a regression branch. To increase the information exchange between the UAV and satellite features in the classification branch, we employ group convolution on the UAV feature map U1 and the current feature map to obtain a response map, which is then upsampled to the original image size using nearest neighbor interpolation.

To balance the positive and negative samples, the authors of WAMF-FPI\cite{bib20} employed a strategy wherein a rectangle with a side length of 33 pixels, centered at the true position in the image, was created. All pixels within the rectangle are treated as positive samples, while the rest are treated as negative samples. It is crucial to have an appropriate number of positive samples for training. However, for the task of finding points with images, what we actually need to find is the point closest to the true position, rather than a range represented by a rectangle. To address this issue, \cite{bib20} introduced the Hanning loss, which assigns different weights to positive samples from different regions. Building upon this, we introduce offset prediction branch in our proposed approach, as shown in Figure \ref{fig_3}. In addition to the classification task, we add an offset prediction taskas a more fine-grained adjustment method, which further improves the localization performance of the model. More details will be discussed in Section \ref{3.4}. The regression branch generates a feature map with 2 channels, where each pixel has two adjustment parameters for modifying the offset in the x and y directions. In Section \ref{5.4}, we conducted a large number of experiments, and the results demonstrate that with the assistance of offset prediction, the proposed model achieves better localization accuracy.

\subsection{Offset Prediction}
\label{3.4}
\begin{figure*}[!t]
	\centering
	\includegraphics[width=6 in]{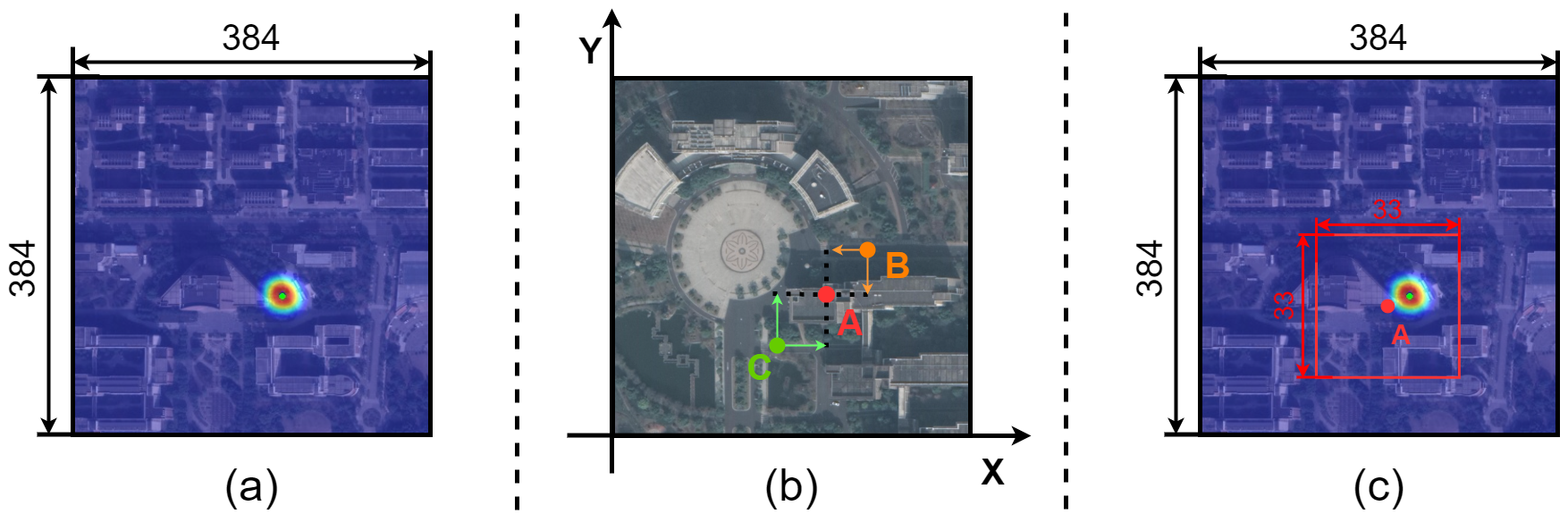}
	\caption{(a) The heatmap outputted by the model in the classification task, where the point with the largest value represents the predicted location of the UAV. (b) Diagram illustrating the adjustment of offset prediction. Point A represents the true position of the UAV, while points B and C represent the classification prediction results of the model. The arrows represent the adjustment process of offset prediction. (c) Point A represents the true position of the UAV. A sample is considered a positive sample if it meets the following two conditions simultaneously:  1. The sample is located within a rectangle with a size of 33 pixels $\times$ 33 pixels centered at point A.  2. On the heat map, the classification score of this sample is the top 300 of all samples.}
	\label{fig_4}
\end{figure*}

Before the introduction of offset prediction, previous methods, such as FPI\cite{bib19} and WAMF-FPI\cite{bib20}, relied on the point with the largest value in the heat map to determine the location of the center of the UAV image. As shown in the heat map in Figure \ref{fig_4}(a), the point with the largest value on the map is the current UAV position predicted by the model. After that, by calculating the position of the pixel in the satellite image, the current latitude and longitude information of the UAV can be calculated according to the ratio. 
In order to achieve more fine-grained positioning and optimize classification predictions, we added offset predictions to the model. That is, the results are adjusted on the basis of classification predictions. As shown in Figure \ref{fig_3}, a new branch is created in the network after the feature is enhanced by atrous convolution. The number of channels output by the model is adjusted to 2, so each pixel will have two adjustment parameters, which are used to adjust the parameters of the x-axis and the parameters of the y-axis respectively.
As shown in Figure \ref{fig_4}(b), point A is the actual position of the UAV. Assume that both points B and C are the classification prediction results of the model. Then, further optimization of the localization results can be achieved by using the adjustment parameters of the offset prediction branch. Obviously, this is a regression task. In the experiments in Chapter \ref{5} , we also showed the performance changes of the model after introducing offset prediction.

After adding the regression task (offset prediction branch), we need to set the positive and negative samples reasonably. How can a sample be considered a positive sample? It needs to meet two conditions. First, as shown in Figure \ref{fig_4}(c), the heat map for classification prediction is shown. A rectangle with a size of 33 pixels $\times$ 33 pixels is drawn centered on point A, where point A is the real position of the UAV, that is, the position of the label. When the sample falls within this rectangular area, it can be considered a positive sample, and if it exceeds this range, it will be treated as a negative sample. Secondly, it must also be satisfied that on the heat map, the classification score of this sample is the top 300 of all samples. Only samples that meet both conditions will be set as positive samples, and the rest of the samples will be ignored. In terms of loss function, we use \begin{math}smooth_{L_{1}}\end{math} \cite{bib44} as the loss function of offset prediction. The formula is as follows: 

\begin{equation}
    smooth_{L_{1}}=
    \begin{cases}
0.5x^2  ~~~~~~~~~ if|x|< 1\\
|x|-0.5 ~~~~~ otherwise \\
\end{cases}
\end{equation}

where x is the difference between the predicted adjustment parameter and the actual adjustment parameter, and Smooth L1 Loss improves the zero point non-smoothness problem compared to L1 Loss. Compared to L2 Loss , when x is large, it is not as sensitive to outliers as L2 Loss , and it is a slowly changing loss function. So the offset loss of the model is: \begin{math}L_{offset}=smooth_{L_{1}}\end{math} 

In addition to the offset loss, OS-FPI also contains the original classification loss, and we follow the Hanning loss in WAMF-FPI in the classification loss part. The Hanning loss\cite{bib20} assumes that the importance of positive samples from different regions is different, so it assigns different weights to positive samples from different regions through the Hanning window function. Equation \ref{eq4} represents the Hanning window function. Therefore the classification loss of the model is: \begin{math}L_{classification}=Hanning \quad loss\end{math}

\begin{equation}
	\label{eq4}
	Hanning(n)=
	\begin{cases}
		0.5-0.5\cos(\frac{2\pi n}{M-1}),0\leq n \leq M-1\\
		0, else\\
	\end{cases}
\end{equation} 

The final loss function formula is as follows:
\begin{equation}
    LOSS= L_{classification} + L_{offset}
\end{equation}

\section{Experiments}
\subsection{Implementation Details}
We trained the OS-FPI on the UL 14 dataset. Our model are implemented using Python 3.7 and PyTorch 1.10.2. The training of the model is conducted on a 1080Ti. Satellite images and UAV images are resized to 384$\times$384$\times$3 and 96$\times$96$\times$3, respectively, with a batch size of 16. We use AdamW optimizer with learning rate of 0.0003 based on cosine scheduling. The learning rate will slowly decrease from 0.0003 to 0.000005. In addition, we set the learning rate of the models other than the backbone to 1.5 times that of the backbone during the training process.

\begin{figure*}[!t]
	\centering
	\includegraphics[width=7 in]{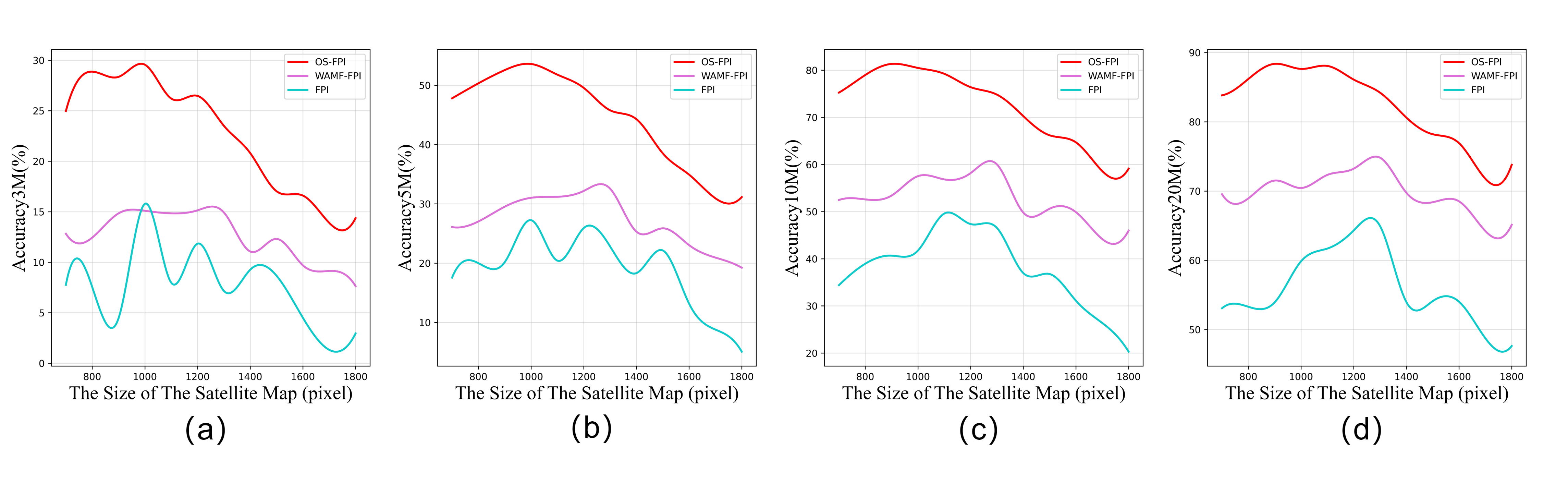}
	\caption{The performance comparison of different models at different scales.}
	\label{fig_6}
\end{figure*}

\begin{figure}[!t]
	\centering
	\includegraphics[width=3 in]{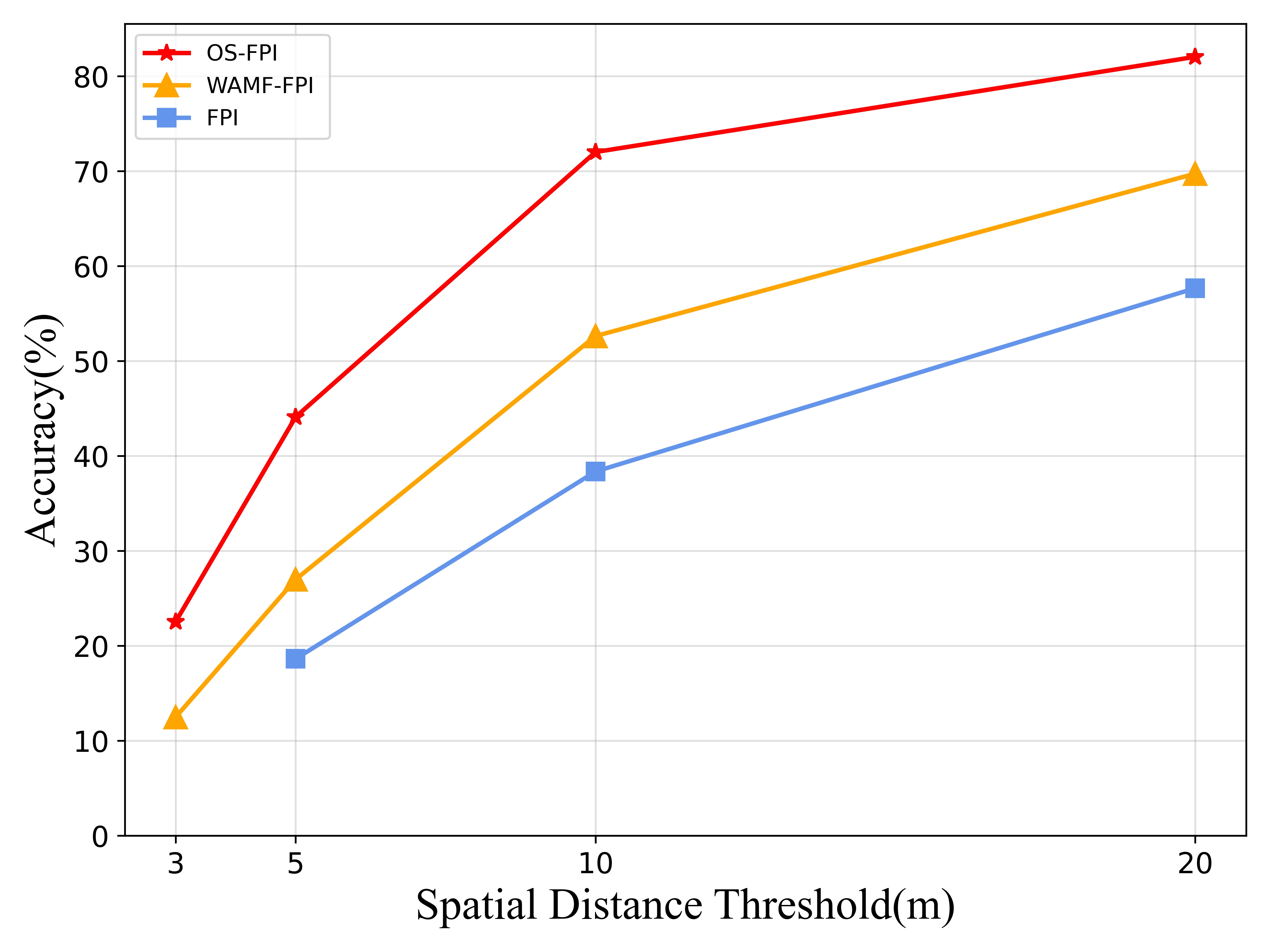}
	\caption{The performance comparison of different models in terms of the MA metric.}
	\label{fig_5}
\end{figure}

\subsection{Dataset and Evaluation Metrics}
\noindent {\bf{Dataset:}}UL14 contains UAV and satellite images of 14 universities in Hangzhou. The UAV images were taken by DJI UAVs at altitudes of 80m, 90m and 100m, with a flight distance of 20m. The image taken by the UAV will be rezized to a size of 512 $\times$ 512 $\times$ 3 after center cropping, and then saved in the database. Afterwards, according to the longitude and latitude information stored in the UAV image, the satellite image of the corresponding area can be cut out from the satellite image, and the cut-out satellite image will be rezized into 1280 $\times$ 1280 $\times$ 3, which will also be sent to the database for storage. It is worth mentioning that the center position of the satellite images is aligned with the center position of the UAV images at this time.
UL14 then divided the dataset, in which 6768 UAV images from 10 universities and 6768 corresponding satellite images were used as training sets (approximately 600 UAV images per university). A further 2331 UAV images from four universities will be used as the test set. 

The satellite images in the test set will be cropped to generate 12 satellite images with different coverage (the side length of the area covered by the satellite image is distributed between 180 meters and 463 meters, including 12 different scales in total). That is, a total of 2331 UAV images and 27972 satellite images are included in the test set. In this way, the difficulty of the test set can be increased, which can also verify the model's ability to solve multi-scale problems. We will follow the previous approach to using this dataset.

\noindent {\bf{Evaluation Metrics:}}In previous works, such as FPI\cite{bib19} and WAMF-FPI\cite{bib20}, RDS and MA were used as evaluation metrics for the models. To ensure fairness, we also employ RDS and MA as evaluation indicators for our proposed model. 

RDS is calculated using equation 3. dx and dy are the pixel distance between the actual position and the predicted position, dx is the pixel distance between abscissas, and dy is the pixel distance between ordinates. w is the width of the satellite image, h is the height of the satellite image. k is the adjustment coefficient, which is set at 10 in this paper.	If the pixel distance between the actual position and the predicted position is closer, the RDS score is closer to 1, otherwise, the closer to 0. 

\begin{equation}
\label{eq3}
RDS= e^{-k\times\sqrt{\frac{(\frac{d_{x}}{w})^2+(\frac{d_{y}}{h})^2}{2}}}
\end{equation}

MA calculates the actual distance deviation between the predicted position and the actual position by latitude and longitude, and the unit of this actual deviation is meter. As the test set contains satellite images of varying scales, the model's positioning performance in the real environment can be accurately and visually displayed using the MA index. For example, the positioning accuracy within 5 meters is defined as the percentage of samples whose distance deviation between the predicted position and the actual position is less than 5 meters to all samples.

RDS pays more attention to the pixel distance between the predicted position and the actual position on the satellite image. In general, RDS measures the pixel distance between the model's positioning result and the actual position (the closer the pixel distance between the actual position and the predicted position in the satellite image, the higher the score), while MA is a direct measure of the true distance between the true location and the predicted location.

\subsection{Comparison with the State-of-the-art models}
We compare our proposed OS-FPI method with previous methods, including the original FPI\cite{bib19} and the latest WAMF-FPI\cite{bib20}. Among the three methods, OS-FPI demonstrates a superior performance. As shown in Figure \ref{fig_5}, the comparison between OS-FPI and previous models under the MA metric reveals that OS-FPI outperforms the other methods. 
Compared to WAMF-FPI\cite{bib20}, our proposed model demonstrates a significant improvement in distribution at distances of 3 meters, 5 meters, 10 meters, and 20 meters, with improvements of approximately 10\%, 17\%, 20\%, and 13\%, respectively. In particular, the performance of the two indicators at 5 meters and 10 meters has been greatly improved.	It is worth mentioning that OS-FPI has higher performance than previous models, but it has higher efficiency and less computational cost. We will compare and analyze this in detail in Section \ref{4.4}.

We also evaluated the performance indicators of the model at different scales. Figures \ref{fig_6} (a), \ref{fig_6} (b), \ref{fig_6} (c), and \ref{fig_6} (d) respectively illustrate the positioning accuracy of the model within distances of 3 meters, 5 meters, 10 meters, and 20 meters. The performance of the original FPI\cite{bib19} model is represented by the light blue line, the purple line represents the performance of the WAMF-FPI\cite{bib20} model, and the red line represents the performance of our proposed OS-FPI model. It can be seen from the figure that OS-FPI has a greater performance improvement compared to other models.

\added{In VIGOR\cite{bib53}, the authors employed image retrieval for UAV geo-localisation. The results of the experiment were obtained through calculations. Their localisation results were less than 10\%, 30\%, and 50\% in 5m, 10m, and 20m, respectively, compared to which OS-FPI demonstrated excellent localisation results.}

\setlength{\tabcolsep}{4pt}
\begingroup
\setlength{\tabcolsep}{6pt} 
\renewcommand{\arraystretch}{1} 
\begin{table*}[!t]
	\begin{center}
		\caption{The detailed comparison between the proposed model and state-of-the-art methods. FPN* denotes the model performance using the feature pyramid structure only in the satellite branches.}
		\label{table1}
		\resizebox{1.0\hsize}{!}{
			\begin{tabular}{cccccccccc}
				\toprule[0.8pt]
				\noalign{\smallskip}
				{\#}&
				{Model}&
				{GFLOPs} &
				{Params}& 
				{Inference Time}&
				{RDS} & 
				{\textless 3m(\%)} &
				{\textless 5m(\%)} &
				{\textless 10m(\%)} &
				{\textless 20m(\%)}
				\\
				\noalign{\smallskip}
				\hline
				\noalign{\smallskip}
				{1} & 
				{OS-FPI}&
				{14.28}& 
				{14.76}& 
				{1.12$\times$}&
				\textbf{76.25} &
				\textbf{22.81} &
				\textbf{44.31} &
				\textbf{72.32} &
				\textbf{82.52}
				
				\\
				{2} & 
				{OS-FPI(FPN*)} & 
				\textbf{10.42}& 
				\textbf{13.84}& 
				\textbf{0.96$\times$}&			
				{66.22} &
				{15.71} &
				{32.51} &
				{57.58} &
				{70.61} 
				\\
				
				{3} & 
				{WAMF-FPI\cite{bib20}} & 
				{13.35}& 
				{48.94}& 
				{1.69$\times$}&
				{65.33} &
				{12.49} &
				{26.99} &
				{52.62} &
				{69.73} 
				\\
				{4} & 
				{FPI\cite{bib19}}&
				{14.88}& 
				{44.48}& 
				{1$\times$}&
				{57.22} &
				{-} &
				{18.63} &
				{38.36} &
				{57.67} 
				\\
				
				\noalign{\smallskip}
				\bottomrule[0.8pt]
			\end{tabular}
		}
	\end{center}
\end{table*}
\setlength{\tabcolsep}{1pt}

\setlength{\tabcolsep}{4pt}
\begingroup
\setlength{\tabcolsep}{6pt} 
\renewcommand{\arraystretch}{1} 
\begin{table*}[!t]
	\begin{center}
		\caption{The performance comparison between a single-stream and two-stream network. GC, FPN*, WAMF indicate the use of different feature fusion methods.
		}
		\label{table2}
		\resizebox{0.9\hsize}{!}{
			\begin{tabular}{ccccccccc}
				\toprule[0.8pt]
				\noalign{\smallskip}
				{\#}&
				{Backbone}&
				{GC} &
				{FPN*}& 
				{WAMF}&
				{OS-FPI} & 
				{RDS} &
				{GFLOPs} &
				{Params} 
				\\
				\noalign{\smallskip}
				\hline
				\noalign{\smallskip}
				{1} & 
				{OS-PCPVT}&
				{\checkmark}& 
				{}& 
				{}&
				{} &
				\textbf{61.28} &
				\textbf{9.91} &
				\textbf{13.47}

				\\
				
				{2} & 
				{PCPVT-S}&
				{\checkmark}& 
				{}& 
				{}&
				{} &
				{56.81} &
				{11.45} &
				{48.21} 
				
				\\
				{3} & 
				{DEIT-S}&
				{\checkmark}& 
				{}& 
				{}&
				{} &
				{55.31} &
				{13.22} &
				{44.42} 
				\\
				{4} & 
				{OS-PCPVT}&
				{}& 
				{\checkmark}& 
				{}&
				{} &
				\textbf{66.22} &
				\textbf{10.42} &
				\textbf{13.84} 
				\\
				
				{5} & 
				{PCPVT-S}&
				{}& 
				{\checkmark}& 
				{}&
				{} &
				{60.02} &
				{12.00} &
				{48.94} 
				\\
				{6} & 
				{OS-PCPVT}&
				{}& 
				{}& 
				{\checkmark}&
				{} &
				\textbf{69.58} &
				\textbf{10.45} &
				\textbf{14.20} 
				\\

				{7} & 
				{PCPVT-S}&
				{}& 
				{}& 
				{\checkmark}&
				{} &
				{64.27} &
				{12.00} &
				{48.94} 
				\\
				
				{8} & 
				{OS-PCPVT}&
				{}& 
				{}& 
				{}&
				{\checkmark} &
				\textbf{76.25} &
				{14.28} &
				{14.76} 
				\\	
				
				\noalign{\smallskip}
				\bottomrule[0.8pt]
			\end{tabular}
		}
	\end{center}
\end{table*}
\setlength{\tabcolsep}{1.4pt}

\subsection{Computational Cost}
\label{4.4}


\added{Table \ref{table1} provides a detailed comparison between the proposed model and state-of-the-art methods. Previous methods used dual branches to extract the features of satellite and UAV images, and because the sources of satellite and UAV images were different, they did not use the method of weight sharing. The resulting problem is a doubling of the number of parameters in the model and a huge computational drain. As shown in Table \ref{table1}, it can be seen that the model after using the one-stream network only supervises the output of the satellite branch and can achieve better positioning results than the previous methods, especially the 3-metre and 5-metre positioning results have a great improvement. It also has less computational complexity and fewer parameters. When more information interactions as well as modules are added to the model, the model improves its RDS score by 19 and 10 compared to FPI and WAMF-FPI, respectively, and at the same time there is a huge improvement in metre-scale positioning accuracy.}

\section{Ablation Experiment }
\label{5}

\subsection{The Effect of One-Stream Structure}

\added{In order to verify the influence of OS-PCPVT on positioning performance after integrating the two functions of feature extraction and information interaction, a series of comparative experiments are presented in Table \ref{table2}. Compared to previous two-stream networks, our method saves a lot of computing resources while allowing better information interaction. In Table \ref{table2}, we show the comparison results between the proposed backbone and the traditional two-stream network. To be fair, all satellite and UAV images are set to 384 $\times 384 \times 3$ and $96 \times 96 \times 3$.}

\added{In ${\#}$1, ${\#}$2, and ${\#}$3, three distinct backbones are employed, and the UAV feature map and satellite feature map output from the last stage of the backbone are directly utilized for relationship modeling, resulting in a response map. After relational modeling, the size of the response map output by the model is 26 $\times$ 26 $\times$ 1. }


\added{The feature pyramid structure was utilized in the experiments of ${\#}$4 and ${\#}$5, but with some variations between them. In ${\#}$4, as feature extraction and information interaction are accomplished simultaneously in the backbone, we only use the feature pyramid network in the satellite image branch. In contrast, ${\#}$5 employs the feature pyramid network in both the UAV and satellite branches, and then relationship modeling is leveraged to facilitate the information interaction between the two branches, resulting in a response map. The results demonstrate that, despite not employing group convolution for information interaction after the backbone, ${\#}$4 achieved superior performance, highlighting the efficacy of the one-stream network in establishing connections between different branches. }

\added{${\#}$6, ${\#}$7 respectively use OS-PCPVT and two PCPVT-S as the backbone, and use the WAMF module as the feature fusion network.}

\added{It can be seen from the three sets of experiments that, under the same conditions, the one-stream network can effectively reduce the number of model parameters and the model complexity. Meanwhile, better results can be obtained. }

\setlength{\tabcolsep}{4pt}
\begingroup
\setlength{\tabcolsep}{6pt} 
\renewcommand{\arraystretch}{1.2} 
\begin{table*}[t]
	\begin{center}
		\caption{The impact of utilizing atrous convolutions with different atrous rates on the performance of the model was investigated, where ${\#}$1 denotes the absence of atrous convolutions, and the numbers in parentheses indicate the corresponding atrous rates.}
		\label{table3}
		\resizebox{1.0\hsize}{!}{
			\begin{tabular}{cccccccccc}
				\toprule[0.8pt]
				\noalign{\smallskip}
				{\#}&
				{Method}&
				{RDS} &
				{\textless 3m(\%)}& 
				{\textless 5m(\%)}&
				{\textless 10m(\%)} & 
				{\textless 20m(\%)} &
				{\textless 30m(\%)} &
				{\textless 40m(\%)} &
				{\textless 50m(\%)}
				\\
				\noalign{\smallskip}
				\hline
				\noalign{\smallskip}
				{1} & 
				{None} &
				{66.22} & 
				{15.71} & 
				{32.51} &
				{57.58} &
				{68.61} &
				{71.35} &
				{73.04} &
				{75.11}
				
				\\
				{2} & 
				{Atrous Convolution(12)}&
				{72.37}& 
				{19.23}& 
				{38.18}&
				{65.96} &
				{78.29} &
				{80.49} &
				{81.52} & 
				{82.95}
				
				\\
				{3} & 
				{Atrous Convolution(12,24)}&
				{72.92}& 
				{19.74}& 
				{38.22}&
				{66.01} &
				{78.49} &
				{81.12} &
				{81.99} & 
				{83.51}					
				\\
				{4} & 
				{Atrous Convolution(12,24,32)}&
				\textbf{73.37}& 
				\textbf{20.18}& 
				\textbf{38.86}&
				\textbf{66.53} &
				\textbf{79.18} &
				\textbf{81.53} &
				\textbf{82.65} & 
				\textbf{84.22}
				\\						
				\noalign{\smallskip}
				\bottomrule[0.8pt]
			\end{tabular}
		}
	\end{center}
\end{table*}
\setlength{\tabcolsep}{1.4pt}

 \begin{figure*}[!t]
	\centering
	\includegraphics[width=7.1 in]{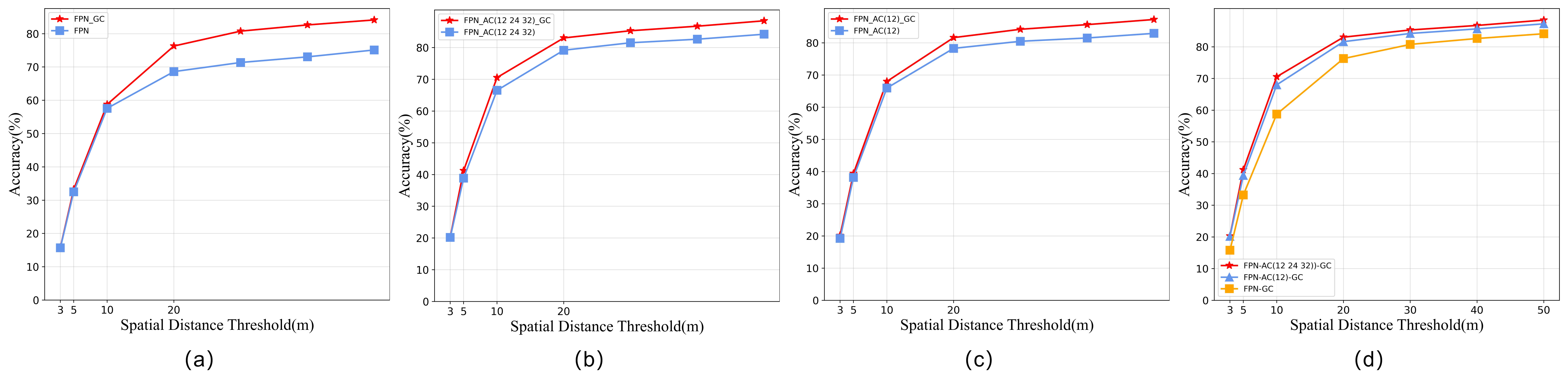}
	\caption{The performance comparison of three control groups in terms of the MA metric, where the horizontal axis represents different spatial ranges.}
	\label{fig_7}
\end{figure*}

\subsection{The Effect of Atrous Convolution}
Through atrous convolution, a larger receptive field can be obtained at a lower computational cost, and more contextual information can be fused at the same time, which is very important for UAV visual positioning. As shown in Table \ref{table3}, we explore different combinations of atrous convolution. We introduced an atrous convolution with an atrous rate of 12 to ${\#}$1. Compared to the original model, the RDS score for ${\#}$2 improved by 6.15, and the accuracy of localisation within 3, 5, 10 and 20 metres improved by 3.52\%, 5.67\%, 8.38\% and 9.68\%. 	Since then, we have added atrous convolutions with atrous rates of 24 and 32. The model's RDS score increased by 0.55 and 1.00, and the positioning accuracy was also improved. In addition, as shown in Figure \ref{fig_7}(d), it also further proves that the introduction of atrous convolution is effective for the UAV visual localisation task.

 \setlength{\tabcolsep}{4pt}
\begingroup
\setlength{\tabcolsep}{6pt} 
\renewcommand{\arraystretch}{1.2} 
\begin{table*}[h]
	\begin{center}
		\caption{ ${\#}$1 represents the localization performance achieved solely by using the classification task, ${\#}$2 represents the localization performance achieved using the classification results after joint training of classification and regression, and ${\#}$3 represents the performance after adjusting the result with the parameters from offset prediction..}
		\label{table5}
		\resizebox{1.0\hsize}{!}{
			\begin{tabular}{cccccccccccc}
				\toprule[0.8pt]
				\noalign{\smallskip}
				{\#}&
				{Method}&
				{GFLOPs} &
				{Params}& 
				{RDS} &
				{\textless 3m(\%)}& 
				{\textless 5m(\%)}&
				{\textless 10m(\%)} & 
				{\textless 20m(\%)} &
				{\textless 30m(\%)} &
				{\textless 40m(\%)} &
				{\textless 50m(\%)}
				\\
				\noalign{\smallskip}
				\hline
				\noalign{\smallskip}
				{1} & 
				{Classification} & 14.1 & 	14.74 & 75.29 & 20.41 & 39.22 &	67.85 &	81.32 &	{83.89} &{85.47} &	{87.22}
				
				\\
				{2} & 
				{Regression and Classification(Classification)} & 14.28 &	14.76 &	75.82 &	21.97 &	41.92 &	69.73 &	82.34 &	\textbf{84.43} &	\textbf{85.67} &	\textbf{87.24}
				\\
				{3} & 
				{Regression and Classification}& 14.28 &	14.76 &	\textbf{76.25} & \textbf{22.81} &	\textbf{44.31} & \textbf{72.32} &	\textbf{82.52} & 84.31 & 85.54 &	87.20
				\\	
				\noalign{\smallskip}
				\bottomrule[0.8pt]
			\end{tabular}
		}
	\end{center}
\end{table*}
\setlength{\tabcolsep}{1.4pt}

\subsection{The Effect of More Information Interaction}
Using a one-stream OS-PCPVT network allows UAV and satellite images to interact with each other in the early feature extraction stage, so does adding more information interaction between UAV and satellite branches after the backbone allow the model to produce better result? To test this idea, we conducted a series of experiments. In order to make a clearer comparison of the improvement in the localization ability of the models. Figure \ref{fig_7} shows the comparison of the three control groups under the MA metric. It can be seen that establishing more information interactions after the backbone can improve the positioning accuracy of the model more substantially at 10m, 20m and beyond. On the basis of the model shown in Figure \ref{fig_2}, we removed the regression branch. GC in the figure indicates that the relationship between the UAV and the satellite feature map is modeled using group convolution, and AC represents atrous convolution using different atrous rates. FPN stands for feature pyramid network, which aims to fuse feature maps of varying scales after the backbone, using a feature enhancement mechanism. We only use the feature pyramid network in the satellite branch.

From the results of the experimental comparison, we found that establishing more information interactions after the backbone helped the model achieve higher RDS scores and higher localization accuracy. The RDS scores of the three control groups increased by 4.12, 3.37 and 3.12, respectively, and at the same time improved in 3 meters, 5 meters, 10 meters and other indicators.


\begin{figure*}[!t]
\centering
\includegraphics[width=6.5 in]{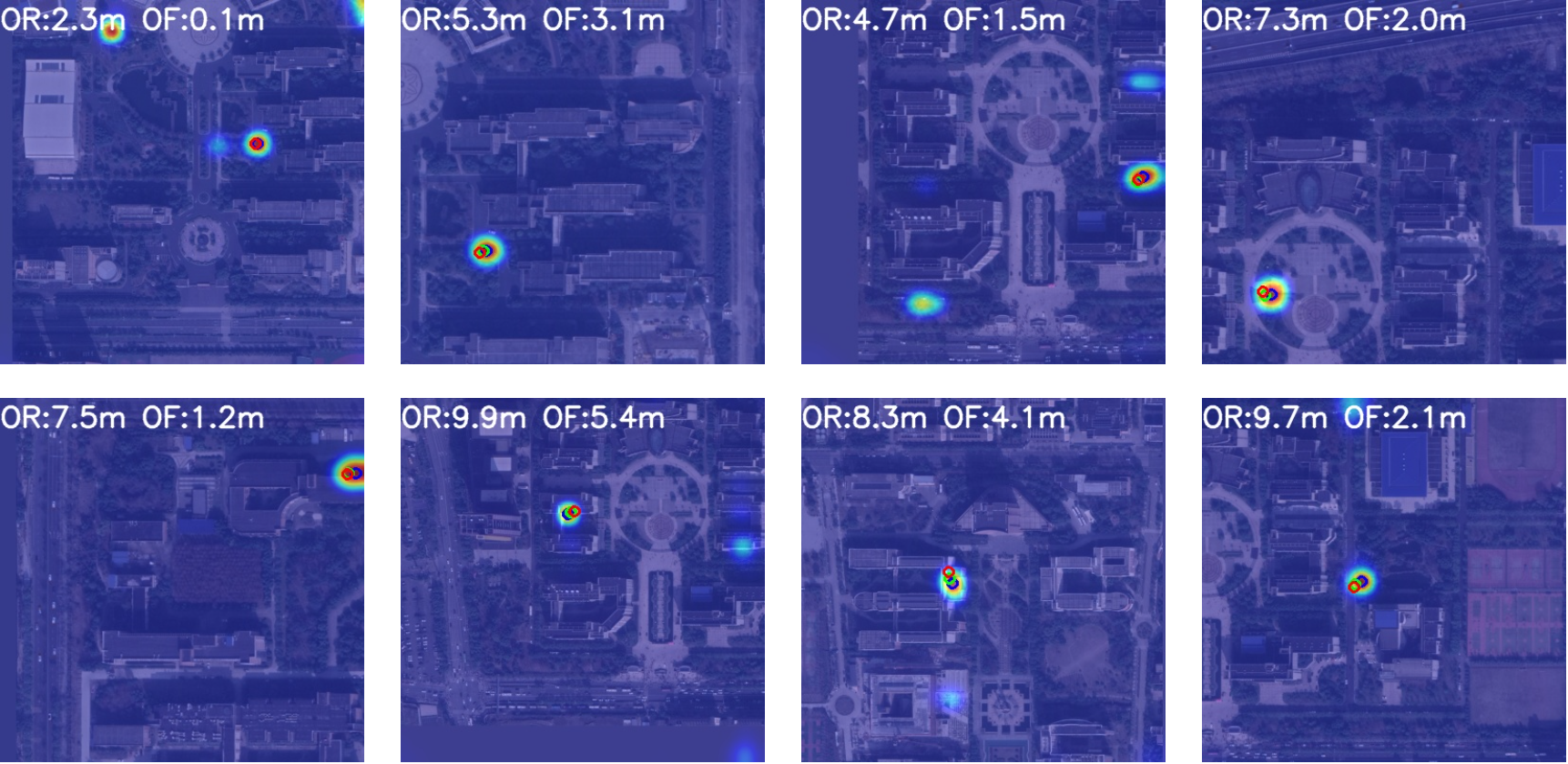}
\caption{The demonstration of the localization performance of OS-FPI, where the heatmap is the classification result of the model, and the point with the largest value in the heatmap is taken as the classification prediction. The red circle represents the true position of the UAV, the blue circle represents the classification prediction, and the green circle represents the optimized result obtained by using offset prediction on top of the classification result. OR represents the localization error between the classification result and the true position, and OF represents the localization error after adjusting the result with the parameters from offset prediction.}
\label{fig_8}
\end{figure*}

%

\subsection{The Effect of Offset Prediction Branch}
\label{5.4}
In this section, we examine the classification branch and the regression branch of the proposed method. In OS-FPI we introduce offset prediction, which means joint training with classification and regression tasks. As shown in Table \ref{table5}, ${\#}$1 indicates the results of training using only the classification branch, ${\#}$2 indicates the prediction results of the classification branch after training using both the classification branch and the offset prediction branch, and ${\#}$3 represents the results of the offset prediction adjustment after using the joint training.

We compared the performance differences between ${\#}$1 and ${\#}$3, and observed that the model with offset prediction achieved an increase of 2.4\%, 5.09\%, and 4.47\% in positioning accuracy within 3, 5, and 10 meters, respectively, as per the experimental results. Furthermore, upon comparing ${\#}$1 and ${\#}$2, it is evident that the introduction of the offset prediction branch, followed by joint training, can enhance the localization performance of the classification branch.

As shown in Figure \ref{fig_8}, it shows the difference in positioning performance before and after the model introduces offset prediction. The heat map in the figure is the result of the classification branch. We take the point with the largest value on the heat map as the result of classification prediction. The red circle represent the actual position of the UAV, the circle dots are the classification predicted positions, and the green circle represent the results of the offset prediction adjustment after using joint training. It can be seen from the figure that the result of the offset prediction can be adjusted across pixels based on the classification prediction, so that the positioning of the model is more accurate. It can be said that this is a more fine-grained positioning method.

However, when comparing the data from ${\#}$2 and ${\#}$3, it is evident that the model's performance experienced a slight decrease after 30m. Moreover, there was minimal improvement in performance when compared to ${\#}$1. The reason for this phenomenon is due to the positive sample setting. During the training process, only the samples within 33$\times$33 pixels centred on the target position can be identified as positive samples. The width and height of the input satellite image are both 384 pixels. And the maximum coverage of the satellite images in the dataset is 463m, so it can be calculated that the coverage of the positive sample is from 0m to 39.78m. Only positive samples in this range are subject to the loss calculation, which is why the accuracy of the model decreases after 30m. Nevertheless, considering the notable enhancement in model performance within the initial 30 meters, we perceive the modest decline in performance beyond that as being reasonable and acceptable.

\begin{figure*}[!t]
	\centering
	\includegraphics[width=6.5 in]{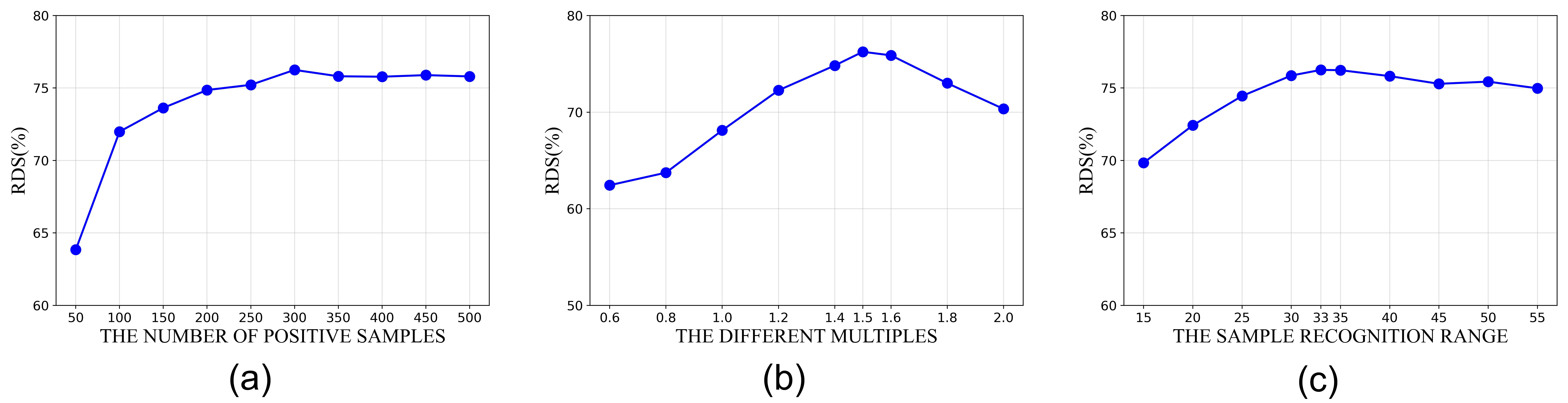}
	\caption{The demonstration of the localization performance of OS-FPI, where the heatmap is the classification result of the model,.}
	\label{fig_20}
\end{figure*}

\begin{figure}[!t]
	\centering
	\includegraphics[width=3.5 in]{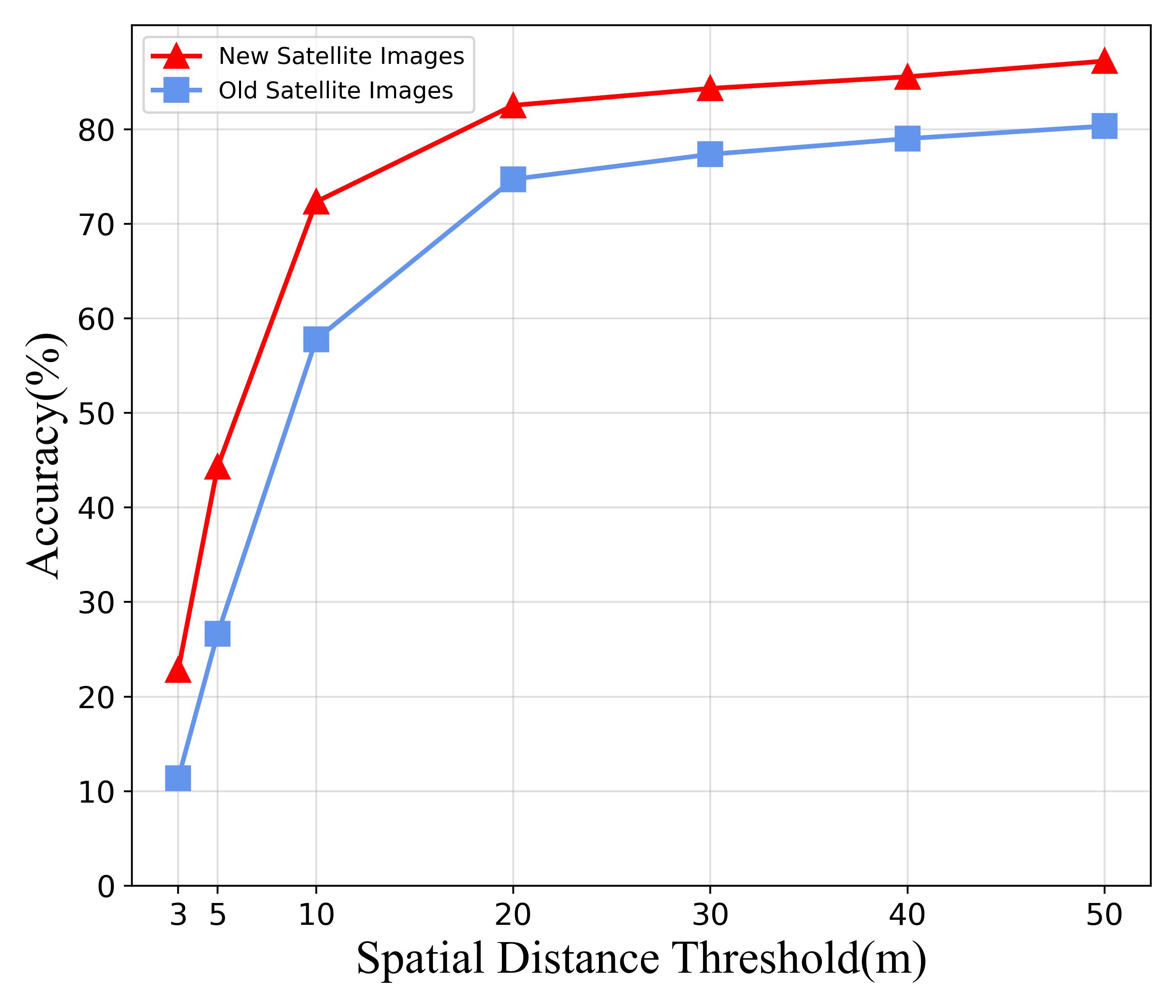}
	\caption{Satellite Maps from Different Time Periods.}
	\label{fig_14}
\end{figure}

\subsection{The Effect of Satellite Maps from Different Time Periods}
\label{5.5}
\added{The infrastructure on the ground may continue to change over time. To ensure the model's robustness in practical applications, we verified its performance using satellite images from different periods. The ground buildings in these images underwent significant changes, posing challenges to accurate positioning of the model. Figure \ref{fig_14} shows the use of satellite images from different periods as the search area. It is evident that changes in the ground infrastructure significantly affect the model's localisation results. Therefore, in future practical applications, the application algorithm should fully consider the impact of time on model performance and adjust the model search range based on flight data and other prior knowledge to enhance practical application ability.}


\subsection{The Effect of Positive Samples}
\label{5.6}
It is widely recognised that the quantity and selection of positive and negative samples greatly impact the training of the model. The number of positive samples, in particular, can have a direct and significant effect on model performance. Therefore, careful consideration should be given to selecting appropriate samples to ensure optimal model training. In OS-FPI, we present a regression branch in which solely positive samples will join the loss computation. Figure \ref{fig_20}(a) illustrates the effect of different numbers of positive samples on model performance. It is evident that the model's performance continuously improves with an increase in positive samples. The findings reveal that selecting the adequate number of positive samples and coverage is imperative for the model's enhanced performance. The results of the experiments show that choosing the right number of positive samples is very crucial for the improvement of the model performance.


\subsection{The Effect of Learning Rate}
\label{5.7}
During training, as the backbone includes pre-trained weights, we consider it essential to distinguish it from other parts and assign distinct learning rates to each part.  As shown in Figure \ref{fig_20}(b), different learning rates are assigned to the rest. Such as, 2 denotes that the learning rate of the other parts is twice that of the backbone. The results of the experiments justify this idea, and assigning a larger learning rate to the parts that do not have pre-trained weights will improve the performance of the model.

\subsection{The Effect of Positive Sample Recognition Range}
\label{5.8}
In order to investigate the effect of the positive sample recognition range in the offset prediction branch, we conducted experiments as shown in Figure \ref{fig_20}(c). The results demonstrate that increasing the range of positive samples within a certain limit enhances the model's localisation performance. However, blindly expanding this range will result in performance degradation.


\begin{figure*}[!t]
	\centering
	\includegraphics[width=6.8 in]{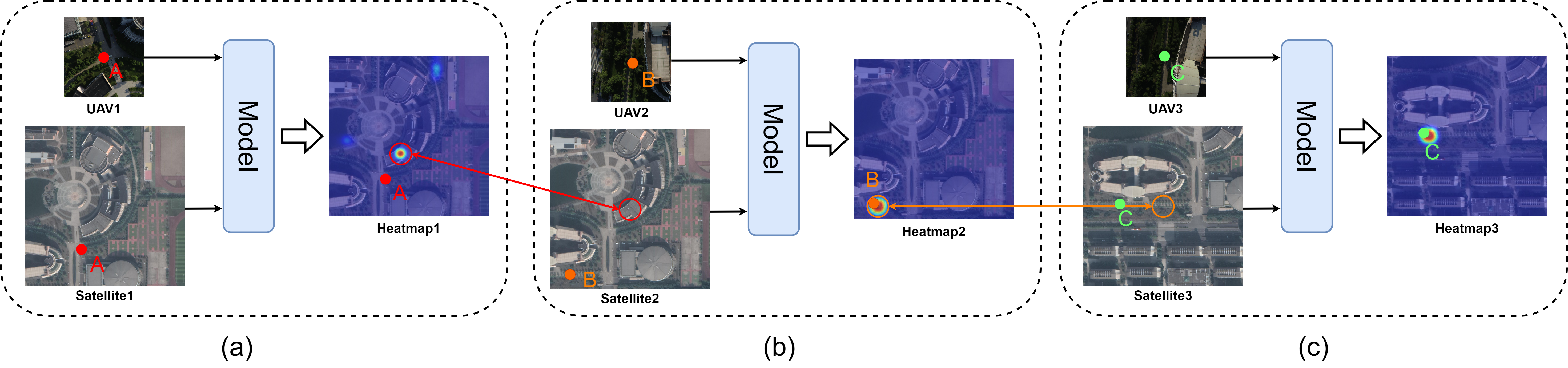}
	\caption{Taking Figure (a) as an example, the red dot A represents the true position of the center of the UAV. After feeding UAV1 and Satellite1 into the network, the corresponding prediction results (heatmap) can be obtained. Then, based on the latitude and longitude of the point with the largest value in the obtained heatmap, a new search area is re-cropped from the satellite image to serve as the search area for the next frame of UAV image.}
	\label{fig_9}
\end{figure*}

\begin{figure}[h]
	\centering
	\includegraphics[width=2.8 in]{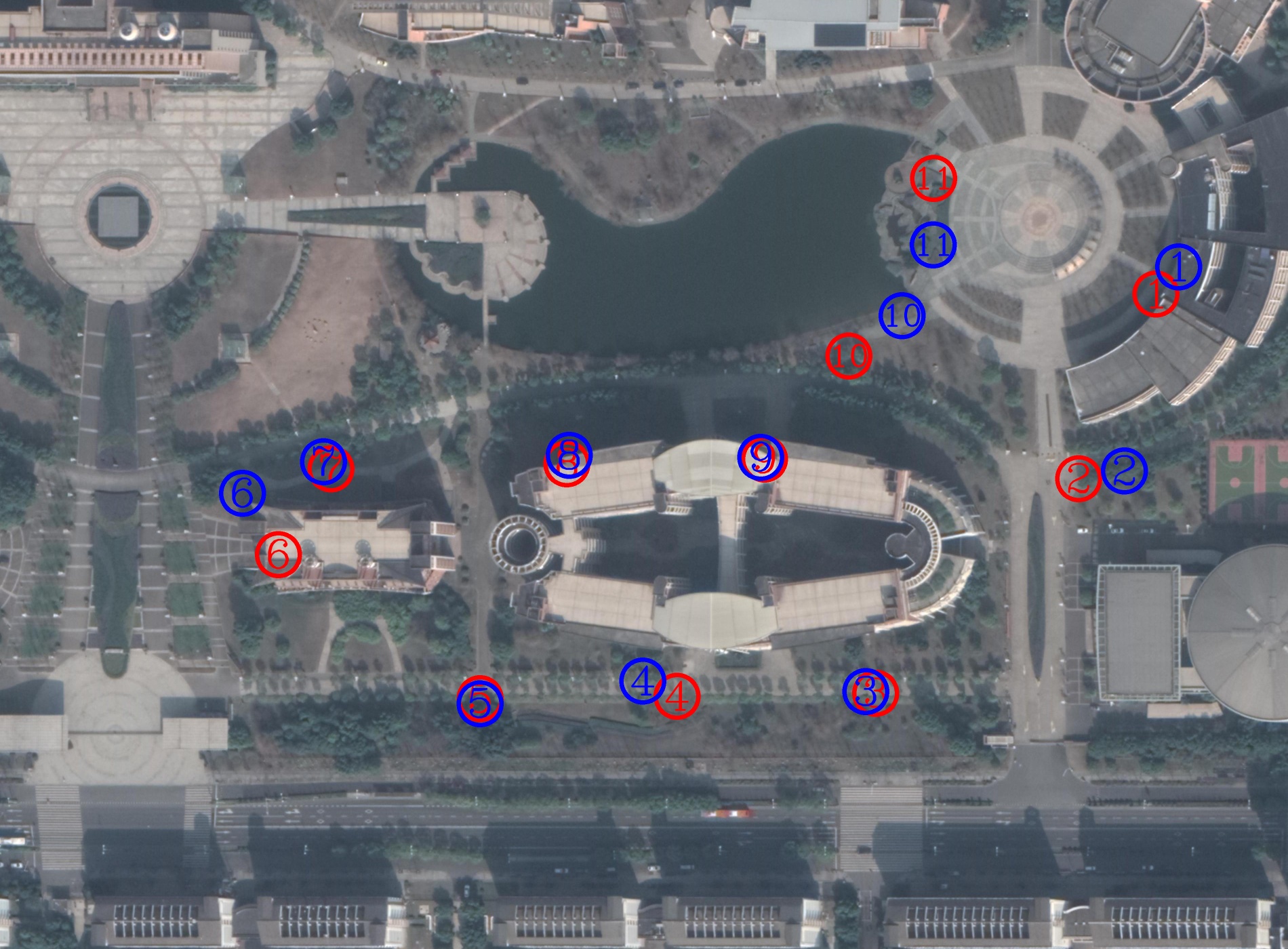}
	\caption{The demonstration of OS-FPI localization performance, where the red label indicates the true position of the UAV and the blue label represents the predicted position by the model.}
	\label{fig_10}
\end{figure}

\section{Application: Assistive Navigation}

After achieving cross-view geolocation, it must be used for navigation in denial environments to truly realize the value of the task. We envision an application scenario where a UAV may lose satellite signals during flight, at which point our algorithm can be used as an auxiliary positioning device to guide the UAV to continue its mission. In order to verify the practical performance of OS-FPI, we conducted a simple application experiment. As shown in Figure \ref{fig_9}, this is a flowchart of the application experiment. First, we will capture an initial search area in the complete satellite map based on the approximate location of where the UAV is located after takeoff. In a real-world environment, this initial position could be the location at which the drone has lost the signal from the satellite. As shown in Figure \ref{fig_9}(a), Satellite1 is cropped in the satellite image according to the approximate position where the first frame of the UAV image is located. Thereafter, UAV1 and Satellite1 are fed into the model at the same time to obtain Heatmap1. The point with the largest heat value in Heatmap1 is the position of the UAV in the satellite image predicted by the model (the offset prediction can also be applied during subsequent experiments), and the latitude and longitude information predicted by the model can be obtained by conversion. Then take the position predicted by the model as the center, re-cut in the satellite image, and obtain the satellite2 (Satellite2 is the search area for the next frame of the UAV image). As shown in Figure \ref{fig_9}(b), the latest position information can be obtained by sending the second frame of UAV images UAV2 and Satellite2 into the model.

Thereafter, the continuous cycle can realize the positioning and navigation of the UAV in the denial environment. As shown in Figure \ref{fig_10}, it shows the positioning effect of OS-FPI. The red label is the actual position, and the blue label is the position predicted by the model. It can be seen that OS-FPI has been able to achieve a certain positioning function, but it needs to continue to improve and optimize.

\section{Discussion}

In recent years, visual geo-localization technology has been a hotspot for research, and most traditional methods use image retrieval to experiment with device localisation, but this method is unable to achieve accurate localisation due to the variation in distance between viewpoints. There are also many practical obstacles to this method. (a) All data in the database must be stored locally after feature extraction and then wait for the query image to be matched. (b)If the model is updated, all data must be re-featured. (c) It takes a long time to retrieve the results. (d) The accuracy of the localisation is highly correlated with the data in the database; the denser the collection, the more accurate and time-consuming the localisation of the model. (e) End-to-end positioning is not possible and requires a lot of preparation.

However, all these problems can be solved by the method of FPI, which is the advantage of the FPI method, which gives more accurate localisation results and does not require much preliminary preparation, which is very practical. FPI research is still at an early stage, and the evolution from the two-stream structure of the WAMF-FPI to the one-stream structure of the OS-FPI has brought a huge improvement in the performance of the model, which also offers the possibility of practical application. In addition, we believe that the refinement and addition of UL14 data will also provide a boost to visual location technology.

\section{Conclusion}
In this paper, we propose a novel, simple, and efficient end-to-end framework called OS-FPI. This is a completely new framework for joint feature extraction and relationship modeling. Unlike the previous two-stream network, OS-FPI has established a link between UAV images and satellite images in the backbone. This means that the connection between UAV images and satellite images is established in the early feature extraction process, which facilitates feature extraction efficiency. At the same time, the introduction of offset prediction for the first time allows the model to further improve the positioning performance of the model on the basis of classification tasks and achieve more fine-grained positioning capabilities. Although OS-FPI has achieved excellent results on the UL14 dataset, there is still huge room for development. From the experimental results, OS-FPI 's ability to solve multi-scale problems still needs to be improved. In addition, further optimization of offset prediction is also an important research direction in the future. The results of the current model can reach more than 70\% within 10 meters, but only about 40\% within 5 meters. Therefore, we believe that there is still great room for improvement in the model, and by continuing to optimize the classification branch and the regression prediction branch of the model, we can definitely achieve more accurate positioning results. In the future, we will also focus more on practical applications and continue to expand the dataset to cover more scenarios.

\bibliographystyle{IEEEtran}
\bibliography{IEEEabrv,re}

\newpage

 
\vspace{11pt}




\vfill

\end{document}